----------
X-Sun-Data-Type: default
X-Sun-Data-Description: default
X-Sun-Data-Name: D.tex
X-Sun-Content-Lines: 463

\documentstyle[12pt]{article}

\title
{\bf\large\bf DETERMINING THE ORDER OF $SU(3)$ DECONFINING PHASE TRANSITION}
\author{{\bf\normalsize\bf Zheng-Kun Zhu}\\
{\small\em Department of Physics and Atmospheric Science, Drexel $University
^{\dagger}$}\\
      {\small\em Philadelphia, Pennsyvalnia 19104-9984}\\
{\small\em Center for Theoretical Physics, Massachusetts
Institute of Technology}\\
{\small\em Cambridge, Massachusetts 02139}\\
{\small and}\\
{\bf\normalsize\bf Da Hsuan Feng}\\
{\small\em Department of Physics and Atmospheric Science,
Drexel University}\\
      {\small\em Philadelphia, Pennsyvalnia 19104-9984}\\
{\small\em Physics Division,
Oak Ridge National Laboratory, Oak Ridge, Tennessee 37996}
}
\date{ }

\begin{document}
\maketitle
\begin{abstract}
An effective spin model for the finite temperature
non-abelian lattice gauge theory is derived.
The outcome is the surprising result that only nearest neighbor
coupling survives, thus confirming the well known numerical results
that the deconfining phase transition of the (3+1)-dimensional
$SU(3)$ pure gauge theory is first order.

%\begin{flushleft}
%PACS numbers: 05.50.+q,11.15.Ha
%\end{flushleft}
\end{abstract}

\newpage

A major success of lattice gauge theory (LGT) is
the numerical demonstration of a transition from the low temperature
color confining to the high temperature non-confining
phase\cite{ukawa90}. Determining the order of the transition is a
subtle matter. For the (3+1) dimensional $SU(3)$ finite temperature LGT,
it was confirmed numerically on the $N_{\tau}=4$ lattice ($N_{\tau}$
is the lattice size in the temporal direction) that the transition
is first order\cite{brown88,fukugita89,iw91}. More recent
calculations on the $N_{\tau}=6$\cite{iw91}
also suggest that this is the case. Still,
the definitive statement about the order of phase transition
remains a challenge for the next generation of computers.

Apart from the full numerical determination the order of the finite temperature
phase transition, there exists  in the literature
universality arguments, first suggested by Svetitsky
and Yaffe in 1982\cite{svetitsky82}, which relate the field
theory to a simpler three dimensional effective spin model.
The derivation of the effective spin model is highly complex.
Nevertheless, by conjecturing a short range coupling, it
was shown that the transition is also
first order\cite{fukugita89,petersson91,gupta90}.
Although the conjecture may not be beyond reproach,
these important results certainly lend confidence to the conclusion
that the transition could indeed be first order.
We therefore feel that there remain the urgent task of deriving an
effective spin model. The purpose of this letter is to carry out the
derivation.

Instead of directly studying the (3+1) dimensional
$SU(3)$ finite temperature LGT,
we will begin by studying a generic $(3+1)$-dimensional $SU(N)$ LGT.
The finite temperature behavior can be described by
a partition function defined on a
hypercubic lattice of size $N_{s}^{3}\times N_{\tau}$,
\begin{equation}
          Q = \sum e^{S},
\label{one}
\end{equation}
where $S$ is the Wilson action
\begin{equation}
        S = \beta_{E}\sum_{n} \sum_{\mu<\nu}
 \frac{1}{N} ReTr(U^{\mu}_{n}U^{\nu}_{n+\mu}
 U^{\mu^{+}}_{n+\nu}U^{\nu^{+}}_{n}),
\label{two}
\end{equation}
$\beta_{E}$ is the coupling constant, $n$ and $\mu,\nu$ represent
the space-time coordinate and directions respectively.
Finite temperature is introduced by
imposing a periodic boundary condition in the time direction,
with period $N_{\tau}$.
%\begin{eqnarray}
%         U^{\mu}_{{\bf x},0} &=& U^{\mu}_{{\bf x},\beta}\nonumber
%\end{eqnarray}
Accordingly, the temperature is $\frac{1}{N_{\tau}a}$, where $a$
is the lattice spacing.

%It is well known that the Wilson line is an order parameter
%for the deconfining phase
%transition.
It is well known that the nature of the
deconfining phase transition can straightforwardly be studied
by  constructing an effective theory
(of the Wilson line)  from the partition function in eq.(\ref{one})
with all the link variables
integrated out except those for the Wilson lines.
Then the effective action $S^{W}_{eff}$
has the following form\cite{svetitsky86}:
\begin{equation}
       S^{W}_{eff} = S^{W}_{eff}(\{\frac{1}{N}TrW_{\bf n}\}).
\label{three}
\end{equation}
Unfortunately, this process
could only be carried out in the strong coupling limit.

There are two main obstacles to obtain $S^{W}_{eff}$.
The first is that in eq.(\ref{two})
there is no explicit Wilson line variable
in $S$. The second is that it is very
complicated to carry out the integration. We will now discuss
the removal of these difficulties.

It turns out that by taking
a thermal gauge choice\cite{creutz77}
the first difficulty can
easily be removed.
\begin{eqnarray}
         U^{\tau}_{{\bf n},n_{\tau}} &=& {\bf 1}  \mbox{    } (1\leq n_{\tau}
  \leq N_{\tau}-1)\nonumber
\end{eqnarray}
The link $U^{\tau}_{{\bf n},N_{\tau}}$ remains unchanged,
and the trace of $U^{\tau}_{{\bf n},N_{\tau}}$
becomes a
Wilson line,
relabelled here as $W_{{\bf n}}$. By inserting the above gauge
choice into eq.(\ref{two}), we can rewrite the action
$S$ of eq.(\ref{two}) as a sum
of $S^{g}$ and $S^{\tau}$, where
\begin{eqnarray}
      S^{g} &=& \beta_{E}\sum_{{\bf n},i<j} \sum_{n_{\tau}=1}^{N_{\tau}}
 \frac{1}{N} ReTr(U^{i}_{{\bf n},n_{\tau}}U^{j}_{{\bf n}+i,n_{\tau}}
 U^{i^{+}}_{{\bf n}+j,n_{\tau}}U^{j^{+}}_{{\bf n},n_{\tau}})\nonumber\\
          & & +\beta_{E} \sum_{{\bf n},i}\sum_{n_{\tau}=1}^{N_{\tau}-1}
 \frac{1}{N} ReTr(U^{i}_{{\bf n},n_{\tau}}
 U^{i^{+}}_{{\bf n},n_{\tau}+1})
\label{four}
\end{eqnarray}
and
\begin{eqnarray}
        S^{\tau}  &=& \beta_{E} \sum_{{\bf n},i}
  \frac{1}{N} ReTr(U^{i}_{{\bf n},N_{\tau}}W_{{\bf n}+i}U^{i^{+}}_{{\bf n},1}
  W^{+}_{\bf n}).
\label{five}
\end{eqnarray}

Straightforward removal of the second difficulty is arduous. To this end, we
will instead effectively decouple
the partition function into two independent sub-partition functions;
one describes the Wilson line field and the other the
space-like link field. We will then derive
the effective spin model by means of a variational principle.
In the following, we will show how to decouple the partition function.

First we will formally write the effective action
as a sum of two parts:
one is an effective spin model, the other has only
space-like link variables\cite{note1}.
\begin{eqnarray}
       S_{eff} \equiv S^{W}_{eff}(\{W_{{\bf n}}\}) +
S_{eff}^{U}(\{U_{n}^{i}\}).
\label{six}
\end{eqnarray}
where $S^{W}_{eff}$ and $S^{U}_{eff}$ are
\begin{equation}
       S^{W}_{eff}=
ln(\frac{\sum_{\{U_{n}^{i}\}}exp(S^{g}+S^{\tau})}
         {\sum_{\{U_{n}^{i}\}}exp(S_{eff}^{U})}),
\label{eq7}
\end{equation}
\begin{equation}
       S^{U}_{eff}=
ln(\frac{\sum_{\{W\}}exp(S^{g}+S^{\tau})}
         {\sum_{\{W\}}exp(S_{eff}^{W})}).
\label{eight}
\end{equation}
We can now see that the partition function of the effective action
given by eq.(\ref{six}) is the
same as that of the action given by eqs.(\ref{four}) and (\ref{five}),
\begin{equation}
        \sum_{U,W}e^{S_{eff}}=\sum_{U,W}e^{S^{g}+S^{\tau}}.
\label{seven}
\end{equation}
In this way at least the decoupling of the fields $W$ and $U$ can
formally be achieved. We can now see that $W$ and $U$ fields can be
described exactly by $S_{eff}^{W}$ and $S_{eff}^{U}$ respectively. It should
be noted that $S_{eff}$ cannot describe the correlation between $W$ and
$U$. However, such a correlation is irrelavant for the physics
discussed in this letter.

Of course, $S_{eff}^{W}$ can in principle
be obtained by intergating ${U}$ in eq.(\ref{eq7}). However as we have
discussed
earlier, it is highly complex to obtain $S_{eff}^{W}$ by this approach.
Instead we have resorted to the variational principle
to derive $S^{W}_{eff}$.
Since $S^{U}_{eff}$ does not explicitly connect
with the deconfining transition in our scheme,
we will not comment on it further. Indeed, there is no loss of generality if
we simply assume that it is already known.

The action
$S_{eff}$ resembles $S$ except for
the coupling between
$W$ and $U$ fields. Therefore it is quite reliable to determine
$S^{W}_{eff}$ by a variational method\cite{drouffe83} where $S_{eff}$ is
a trial action. To this end,
we will first determine its form.
We notice that there is a local gauge
invariance for $W$ in $S$ of eqs.(\ref{four}) and (\ref{five}).
Hence, in order to maintain this invariance, $S_{eff}^{W}$
 must only depend on $Tr(W^{m})$ (where m is an integer)\cite{ogilvie84}.
Then we can describe the action $S^{W}_{eff}$
as follows:
\begin{equation}
     S^{W}_{eff} = \beta_{E}\alpha\sum_{{\bf n},i}
  Re( \frac{1}{N} TrW_{{\bf n}+i}\frac{1}{N}TrW^{+}_{\bf n})
    + S^{W}_{r}(\{\frac{1}{N}Tr(W_{\bf n}^{m})\}).
\label{nine}
\end{equation}
where $\alpha$ is a variational parameter.
 From eq.(\ref{five}), we note that
the first term of eq.(\ref{nine}) is the dominant part of the effective
spin model and $S^{W}_{r}$ the residue.
What remains is to determine $\alpha$ and the form of $S_{r}^{W}$.

Ignoring the
residue part in eq.(\ref{nine}) for the moment,
we write the action as:
\begin{eqnarray}
     S_{0}^{W} &=& \alpha \beta_{E} \sum_{{\bf n},i}
   Re (\frac{1}{N} TrW_{{\bf n}+i}\frac{1}{N}TrW^{+}_{\bf n}),
\end{eqnarray}
then the trial action becomes
\begin{equation}
     S_{0} = S_{0}^{W} + S^{U}_{eff}.
\end{equation}
To determine the value of $\alpha$, we calculate
the
partition function as follows,
\begin{eqnarray}
     Q &=& Q_{0}<exp\{S^{g}+S^{\tau}-S_{0}\}>_{0},
\end{eqnarray}
where $<\cdots>_{0}$ represents the average in action
$S_{0}$ and $Q_{0}$ is
\begin{equation}
              Q_{0}=\sum e^{S_{0}}.
\end{equation}
Using Jensen's inequality $(<exp X> \geq exp<X>)$, we can obtain
\begin{equation}
       F=lnQ \geq lnQ_{0} +<S^{g}+S^{\tau}-S_{0}>_{0}
\end{equation}
and $\alpha$ can then be determined by maximizing $F_{var}$
with respect to $\alpha$
\begin{equation}
    F_{var}=lnQ_{0}+<S^{g}+S^{\tau}-S_{0}>_{0}.
\end{equation}
To accomplish this, we will compute $<S^{g}+S^{\tau}-S_{0}>_{0}$ (
$\equiv K$):
\begin{eqnarray}
    \frac{K}{N_{s}^{3}}&=&3(\beta_{E}Re(<\frac{1}{N}TrW_{{\bf n}}^{+}
\frac{1}{N}TrW_{{\bf n}+i}>_{W 0} <\frac{1}{N}Tr(U^{i}_{{\bf n},N_{\tau}}
U^{i^{+}}_{{\bf n},1})>_{U})\nonumber\\
                      & &  -\alpha \beta_{E}Re(<\frac{1}{N}TrW_{{\bf n}}^{+}
\frac{1}{N}TrW_{{\bf n}+i}>_{W 0}))\nonumber\\
                     & & + \frac{<S^{g}-S^{U}_{eff}>_{U}}{N_{s}^{3}}
\end{eqnarray}
where $<\cdot\cdot\cdot>_{U}$ and $<\cdot\cdot\cdot>_{W 0}$
represent the average in $S^{U}_{eff}$ and
$S_{0}^{W}$ respectively.
We then notice that $S^{W}_{0}$ is real and
invariant under the transformation
\begin{eqnarray}
     W &\rightarrow& W^{+},
\label{seveteen}
\end{eqnarray}
hence
$<\frac{1}{N}TrW_{{\bf n}}^{+}
\frac{1}{N}TrW_{{\bf n}+i}>_{W 0}$ is also real.
By maximizing $F_{var}$ with respect to $\alpha$
, we obtain
\begin{equation}
     \alpha = Re(<\frac{1}{N}Tr(U^{i}_{{\bf n},N_{\tau}}
U^{i^{+}}_{{\bf n},1})>_{U})
\label{eighteen}
\end{equation}

We are now ready to determine
the form of $S^{W}_{eff}$. Let's first write $S_{r}^{W}$ as
\begin{eqnarray}
      S_{r}^{W} \equiv \xi S_{r}^{W M}
\label{twenty}
\end{eqnarray}
where $\xi$ is a variational parameter and $S_{r}^{W M}$ the remaining
part of the effective action which is independent of $\xi$.

We notice that the action $S$ of eqs.(\ref{four}) and (\ref{five}) is invariant
 if all the link variables $U$ transform as
$U \rightarrow U^{+}$(here $U$ means either $U$ or $W$). To maintain this
 invariance
in the effective spin model,
$S_{eff}^{W}$ must be invariant under the transformation $W \rightarrow W^{+}$.

Since $S_{eff}^{W}$ is real and invariant under $W \rightarrow W^{+}$,
$<\frac{1}{N}TrW_{{\bf n}}^{+}
\frac{1}{N}TrW_{{\bf n}+i}>_{W}$
(where $<\cdot\cdot\cdot>_{W}$
represents the average in action
$S^{W}_{eff}$,) is still real.
With this, we will again compute the partition function
$Q_{eff}$ in the action $S_{eff}$ and $K$ as:
\begin{eqnarray}
    K &=& <S^{g}-S_{eff}^{U}>_{U}
                       -<S^{W}_{r}>_{W},
\end{eqnarray}
By maximizing $F_{var}= ln Q_{eff} + K $ with respect to $\xi$,
we obtain
\begin{equation}
    <\frac{\partial S^{W}_{r}}{\partial \xi}>_{W}
   -\frac{\partial <S^{W}_{r}>_{W}}{\partial \xi} =0.
\label{twentythree}
\end{equation}
 From eqs.(\ref{twenty}) and (\ref{twentythree}),
we obtain
\begin{equation}
     \xi\frac{\partial <S_{r}^{W M}>_{W}}
{\partial \xi}=0.
\label{twentyfour}
\end{equation}
and
\begin{equation}
     \frac{\partial <S_{r}^{W M}>_{W}}
{\partial \xi}=<(S_{r}^{W M})^{2}>_{W}-<S_{r}^{W M}>_{W}^{2}
\label{twentyfive}
\end{equation}
Clearly according to eq.(\ref{twentyfive}), $\frac{\partial <S_{r}^{W M}>_{W}}
{\partial \xi}$ does not vanish unless $S_{r}^{W M}$ is a constant,
a situation which
is not relavant in our discussion. From eq.(\ref{twentyfour}), we see that
$\xi$ must vanish.
Our derivation is general since $S_{r}^{W M}$
includes any kind of coupling terms.
Hence surprisingly, we have obtained an effective spin model with only nearest
 neighbor coupling terms.

It is important to understand why by using the variational principle, the
 effective spin model appears simple. Actually, it is very complex in the sense
 that to compute $\alpha$ according to eq.(\ref{eighteen}) is complicated.
The reason is as follows: In order to obtain $\alpha$, we have to know the form
 of $S_{eff}^{U}$ which can be obtained by integrating over $W$ in
 eq.(\ref{eight}). With this knowledge, $\alpha$ can then be calculated
 according to eq.(\ref{eighteen}).
Both procedures are nontrivial. In fact, the difficulty of deriving the
 effective spin model has actually been transferred from the integration of $U$
 in eq.(\ref{eq7}) to the integration of $W$ in eq.(\ref{eight}) and
to calculate $\alpha$ from eq.(\ref{eighteen}).  Therefore the simplicity of
the
 form of the effective spin model is not obtained without a price. What we have
done is to "organize" the theory such that the discussion of
the order of deconfining phase transition can be vividly studied.

Having the effective spin model, we can now focus on studying the deconfining
 phase transition. To this end, we note that there were numerous investigation
 made in this direction already. These studies enable us to discuss the order
of
 the deconfining phase transition.

For the $SU(2)$ gauge theory, it is well known that the plaquette is continuous
 with the change of $\beta_{E}$. Therefore, according to eq.(\ref{eighteen}),
 $\alpha$ is also a continuous function of $\beta_{E}$. Hence our study
directly
 shows that the phase transition of the $Z(2)$ spin model possesses the same
 universality property as the $SU(2)$ gauge theory. This agrees with the
 recently reached conclusion by the Monte-Carlo real space renormalization
 study\cite{bitar88,note3}.

For the $SU(3)$ gauge theory, it was shown via the Monte Carlo study of the
 effective spin model with a nearest neighbor
coupling assumption that the transition is first
order\cite{ukawa90,gupta90}.
Results reported here confirm the nearest neighbor assumption and
are consistent with the recent Monte-Carlo results about the
transition order\cite{fukugita89,petersson91}.

Finally, we mention
that our derivation appears to provide an analytical approach to
study the thermodynamical behavior of LGT\cite{zhu92}.

We are very grateful to Lay Nam Chang, Xiangdong Ji, Robert Perry,
Hai-Chang Ren and Yong-Shi Wu for illuminating discussions. We also
thank N. Christ, M. Fukugita, S. Ohta and C. DeTar for several useful
communications and suggestions.
One of us (ZKZ) is very grateful for
Center for Theoretical Physcis of MIT for its hospitality.
This work is supported by
the National Science Foundation and the Department of Energy.

\begin{flushleft}
$^{\dagger}$ Permanent address
\end{flushleft}


\begin{thebibliography}{20}
\bibitem[14]{note3} There exist non-nearest coupling terms in the
Potts model, although the range is exponentially short.
Our results are not in contradiction with such results because
we have used the $SU(N)$ matrix and not the $Z(N)$ spin in our
calculations.
\end{thebibliography}
\end{document}